\documentclass[
 reprint, 
 superscriptaddress,
 amsmath,
 amssymb,
 aps, 
 physrev
]{revtex4-2}

\usepackage{graphicx}% Include figure files
\usepackage{dcolumn}% Align table columns on decimal point
\usepackage{bm}% bold math
\begin{document}

%\title{Deep learning-enhanced 3D Tracking of motile microorganisms in complex environments}
%\title{Deep learning-enhanced 3D Tracking of motile microorganisms}
\title{Deep learning-enhanced Lagrangian 3D Tracking of motile microorganisms}

\author{Thierry Darnige}
\affiliation{Laboratoire PMMH-ESPCI Paris, PSL Research University, Sorbonne Universit\'e and Denis Diderot, 7, quai Saint-Bernard, Paris, France.}

\author{Daniel Midtvedt}
\affiliation{Physics Department, University of Gothenburg, Sweden.}

\author{Renaud Baillou}
\affiliation{Laboratoire PMMH-ESPCI Paris, PSL Research University, Sorbonne Universit\'e and Denis Diderot, 7, quai Saint-Bernard, Paris, France.}

\author{Benjamin Perez Estay}
\affiliation{Laboratoire PMMH-ESPCI Paris, PSL Research University, Sorbonne Universit\'e and Denis Diderot, 7, quai Saint-Bernard, Paris, France.}

\author{Changsong Wu}
\affiliation{Laboratoire PMMH-ESPCI Paris, PSL Research University, Sorbonne Universit\'e and Denis Diderot, 7, quai Saint-Bernard, Paris, France.}
\author{Alex Le Guen}
\affiliation{Laboratoire PMMH-ESPCI Paris, PSL Research University, Sorbonne Universit\'e and Denis Diderot, 7, quai Saint-Bernard, Paris, France.}

\author{Giovanni Volpe}
\affiliation{Physics Department, University of Gothenburg, Sweden.}
\affiliation{Science for Life Laboratory, Physics Department, University of Gothenburg, Sweden.}

\author{Eric Cl\'ement}
\affiliation{Institut Universitaire de France (IUF).}
\affiliation{Laboratoire PMMH-ESPCI Paris, PSL Research University, Sorbonne Universit\'e and Denis Diderot, 7, quai Saint-Bernard, Paris, France.}

\date{\today}

\begin{abstract}
How microorganisms respond to and interact with their environment can vary significantly from individual to individual, which can have important microbiological and ecological implications.
However, most microscopy techniques can only observe motile microorganisms for short times because of their limited fields of view.
Using Lagrangian tracking, a single microorganism can be followed in 3D, potentially indefinitely, allowing to decipher individual phenotypical traits. 
Current Lagrangian tracking methods use the fluorescence signal emitted by the microorganism as feedback to keep it in focus.
However, over long times, epifluorescent imaging can induce photobleaching and photodamage, and importantly, not all microorganisms can easily be made fluorescent.
Additionally, traditional algorithms used in feedback loops to determine microorganism position are prone to errors, especially in optically complex media.
Here, we present a faster, more reliable, and versatile Lagrangian tracking method that uses deep learning to determine the 3D position of the microorganism.
This new method demonstrates enhanced accuracy and speed in tracking fluorescent bacteria with fluorescence microscopy also in optically complex media. 
Furthermore, we track bacteria with other microscopy modalities, such as brightfield microscopy---for example, this enables us to track magnetotactic bacteria, which cannot be made fluorescent without degrading their magnetotactic properties. 
These novel capabilities allow  to extract previously inaccessible quantitative information, significantly advancing the study of microorganism behavior---and thus opening new avenues for research in complex biological and ecological systems.
\end{abstract}
\maketitle
\begin{figure*}[ht!]
%\centering
\includegraphics [width=\linewidth]{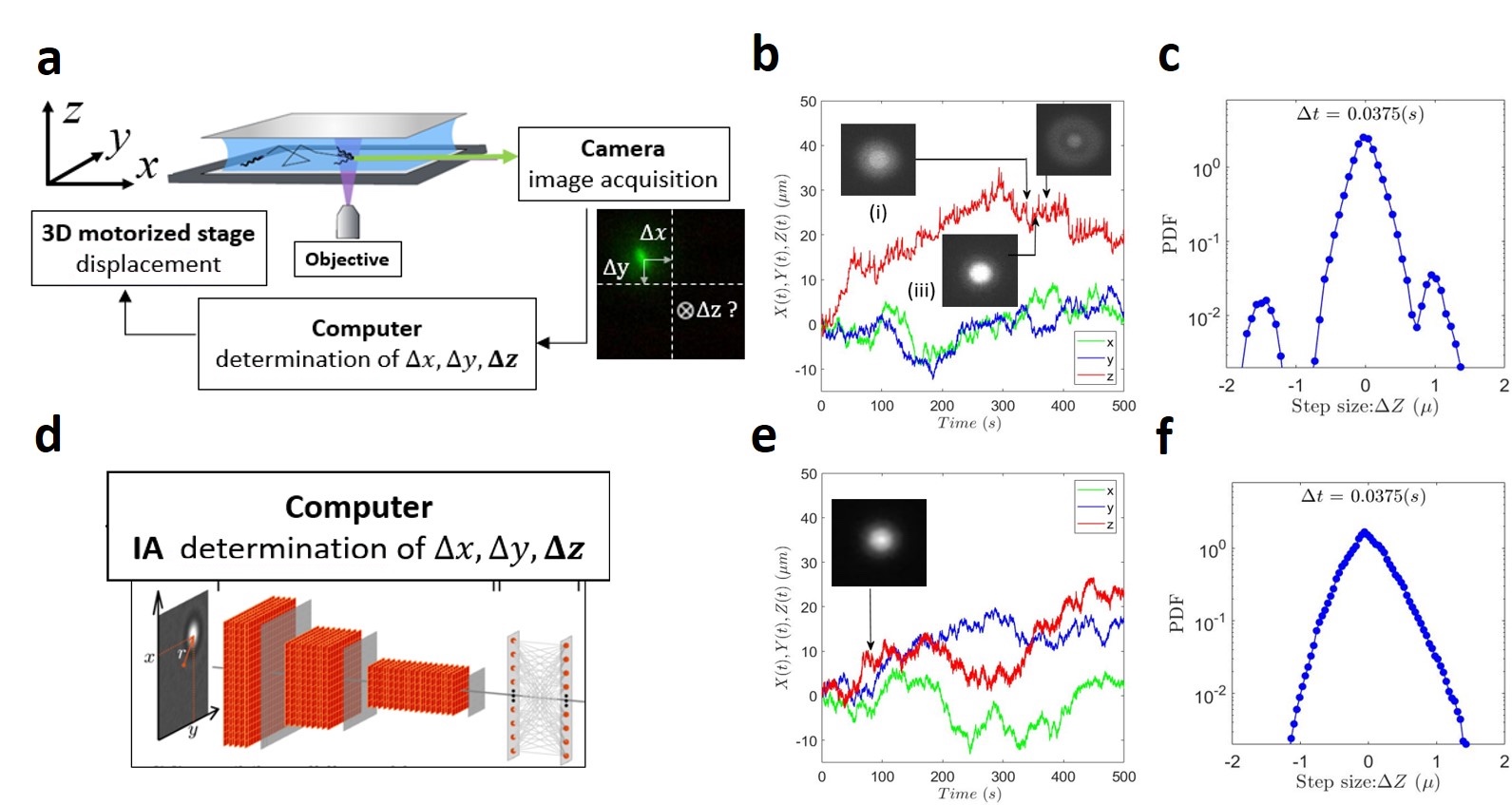}
\caption{
{\bf AI-enhanced Lagrangian tracking.}
(a) Minimalist sketch of a standard Lagrangian tracking setup featuring the feedback loop between the image of a particle  (see Darnige et al. \cite{darnige2017lagrangian} for  details on the algorithm)  and the command on a set of 3 stages moving in $X,Y,Z$ suited to keep the particle in the central part of the camera image and in the focal plane position $z_F$ (More details in  supplementary document Fig S1).
(b) Brownian traces $X(t), Y(t), Z(t)$ of a fluorescent latex bead of diameter $d = 1.1 \pm 0.1 \mu$m obtained with the standard tracking algorithm. Inset images are zooms on the particle that highlights a ``jump'' in $Z(t)$ corresponding to a sudden loss of focus (see Video1b in supplementary documents) . 
%The corresponding particle images on the camera are also displayed.
(c) Statistics of the short-time displacements (time lag $\Delta t = 0.0375 s$) showing the statistical importance of spurious jumps due to tracking errors for the standard set-up. %fat tails, non-Gaussian behavior.
(d) Sketch of the neural network (NN) used in the new 3D-tracking version to compute the expected shifts from the central positions $\Delta X$, $\Delta Y $ and of the focal plane  $\Delta Z $. This new algorithm is used to feedback on the stages positions.
(e) 3D traces of the Brownian motion for a $d = 1.1 \pm 0.1 \mu$m latex bead displaying no spurious ``jump''. Inset image of the bead, essentially  kept in focus. (see Video1e in supplementary documents)
(f) Statistics of the short-time displacements (time lag $\Delta t = 0.0375 s$) showing the absence of spurious "jumps".
}
\label{fig_1} 
\end{figure*}
\section{Introduction}
Understanding how individual microorganisms navigate and interact with their environment is fundamental to microbiology, ecology, and medical sciences\cite{Elgeti_2015}. The behaviors of single cells, such as bacteria and protozoa, can influence biofilm formation, infection processes, and nutrient cycling in ecosystems \cite{Hall_2004,Zheng_2024}. However, capturing the full scope of these behaviors over extended periods, remains a significant challenge due to technical limitations in microscopic observation techniques.

Classical tracking methods under the microscope often provide only fleeting glimpses of motile microorganisms, as they quickly move out of their limited field of view. State to the art  high-throughput 3D tracking methods \cite{Molaei_2014,Taute_2015} have been developed more recently, however the tracking time constraint still limits the ability to study long-term behaviors, heterogeneity among individual cells \cite{Emonet2008}, and rare events that could be crucial for understanding complex biological contamination processes \cite{figueroa2020coli}. Techniques like tethering cells \cite{Korobkova2004}, optically trapping them \cite{mears2014escherichia}, or confining them within narrow microfluidic channels \cite{Kantsler_2013} have been employed, but these methods can alter natural behaviors and interactions.
Lagrangian tracking offers a solution by following individual microorganisms through three-dimensional (3D) space over time, effectively moving the observation frame along with the subject \cite{berg1971track}. A recent implementation of this method utilizes fluorescence microscopy, where the emitted light from labeled microorganisms provides feedback to keep them in focus \cite{darnige2017lagrangian}. While successful in extending observation times, fluorescence-based tracking has inherent drawbacks. Prolonged exposure to excitation light can lead to photobleaching and photodamage, potentially affecting cell viability and behavior \cite{Boudreau_2016}. Moreover, not all microorganisms are amenable to fluorescent labeling without altering their physiological properties---for example, fluorescent tagging expressed at high oxygenation can disrupt the magnetosome formation of magnetotactic bacteria that requires a low oxygen level  \cite{Heyen2003}.

Another significant limitation arises in complex and crowded environments, such as biofilms or natural mucus, where optical distortions and background noise complicate accurate tracking \cite{Urra_2025}. Standard algorithms struggle to maintain focus and positional accuracy in these situations, leading to frequent loss of the tracked microorganism and unreliable data. This gap highlights the need for a more robust, versatile tracking method that can operate effectively across different microscopy modalities and challenging environments.

Here, we present a novel Lagrangian tracking method that uses deep learning to determine the three-dimensional position of individual microorganisms with high accuracy and speed. Our approach eliminates the reliance on fluorescence for feedback, enabling the use of brightfield microscopy and thereby avoiding photobleaching and photodamage. We demonstrate the capability of our method to track motile bacteria, including strains that cannot be fluorescently labeled without compromising their properties, such as magnetotactic bacteria. Additionally, we showcase the method's robustness in optically complex media, allowing for the extraction of quantitative data previously inaccessible with traditional tracking techniques. This new Lagrangian method opens new avenues for studying microorganism behavior in realistic environments, enhancing our understanding of their ecological and physiological roles.

\section{Results}

\subsection{Deep learning enhances Lagrangian tracking accuracy}
In conventional Lagrangian tracking systems, microscopic objects such as microorganisms or particles are kept in focus by adjusting the microscope's position based on feedback from image features like brightness or apparent object size \cite{darnige2017lagrangian}. These standard tracking algorithms often rely on optimizing certain criteria (e.g., maximizing fluorescence intensity or minimizing particle apparent size) to determine the direction and magnitude of the required adjustments. Fig.~\ref{fig_1}(a) illustrates a minimalistic standard setup for Lagrangian tracking, highlighting the feedback loop used to maintain the object in focus. See also Methods ``Experimental setup  and imaging system'' \ref{A} and Supplementary document Fig. S1 for a detailed schematics of the setup. The object’s image is captured by the camera, processed by the feedback algorithm, and used to adjust the position of the microscope stage to keep the object centered and in focus.

These tracking algorithms can struggle with maintaining accurate focus, particularly in the axial z-direction. This is evident when tracking the three-dimensional (3D) Brownian motion of colloidal particles. To illustrate this issue, we use Fluoromax polystyrene spheres of diameter $d=1.1\pm 0.1 \mu$m dispersed in a colloidal dispersion achieving  non-buoyant conditions (Percoll Sigma-P1644 $50 \%$). Fig.~\ref{fig_1}(b) shows 3D traces of a Brownian spherical particle obtained using the standard tracking method. The inset reveals details of ``jumps'' in the z-position, indicating instances where the feedback algorithm failed to maintain smooth tracking due to ambiguous signals near the focal plane.
To quantify the effect of these tracking errors, we analyzed the statistics of the short-time displacements. Fig.~\ref{fig_1}(c) presents the distribution of the particle's short-time displacements in the z-direction, which displays secondary maxima
%and deviations from the expected Gaussian behavior. These non-Gaussian distributions are 
indicative of sporadic errors introduced by the standard feedback algorithm, which can significantly distort the statistical analysis of particle motion or eventually lead to tracking failures.

To overcome these limitations, we developed a deep learning-based tracking algorithm that enhances the accuracy and robustness of the Lagrangian tracking. The new setup is depicted in Fig.~\ref{fig_1}(e), which includes the integration of a convolutional neural network (NN) into the feedback loop (see also Methods ``Deep learning model development'' \ref{D} and Fig. S2 in Supplementary document for the details of the neural network architecture). Based on the current image as input, this neural network is trained to predict the 3D position of the particle, enabling rapid and precise adjustments to the microscope's stage position (see also Fig. S3 and Fig. S4 in Supplementary document for details of the training procedure). 

Using the deep-learning-based feedback, we tracked the same Brownian colloidal particles and observed a significant improvement in tracking performances.

Fig.~\ref{fig_1}(e) shows the 3D traces obtained with the deep-learning feedback algorithm. The inset highlights the absence of ``spurious jumps'' seen with the standard method, indicating smoother and more accurate tracking %The trajectory appears continuous and free from abrupt displacements, demonstrating the algorithm's ability to maintain focus without interruption.
(See also  Fig. S5 in Supplementary document for the corresponding  mean square displacements).
The statistical analysis of the short-time displacements confirms this improvement. As shown in Fig.~\ref{fig_1}(f), the distribution of the particle's short-time displacements in the z-direction displays  no more spurious jumps.
%now closely follows a Gaussian distribution, as expected for Brownian motion.

It is important to realize that the tracking improvement visualized here, results from a paradigmatic shift in the method by which the algorithm determines the feedback information delivered to the stage. While the former algorithm defines the relevant information to move the stage  as reaching an optimum, the deep-learning algorithm builds a \textit{zero-crossing} method to determine whether the visualization plane is above or below the focal position. Therefore, this method is in principle ideal for providing accurate feedback on the stage.

\begin{figure}[t!]
\includegraphics [width=9cm]{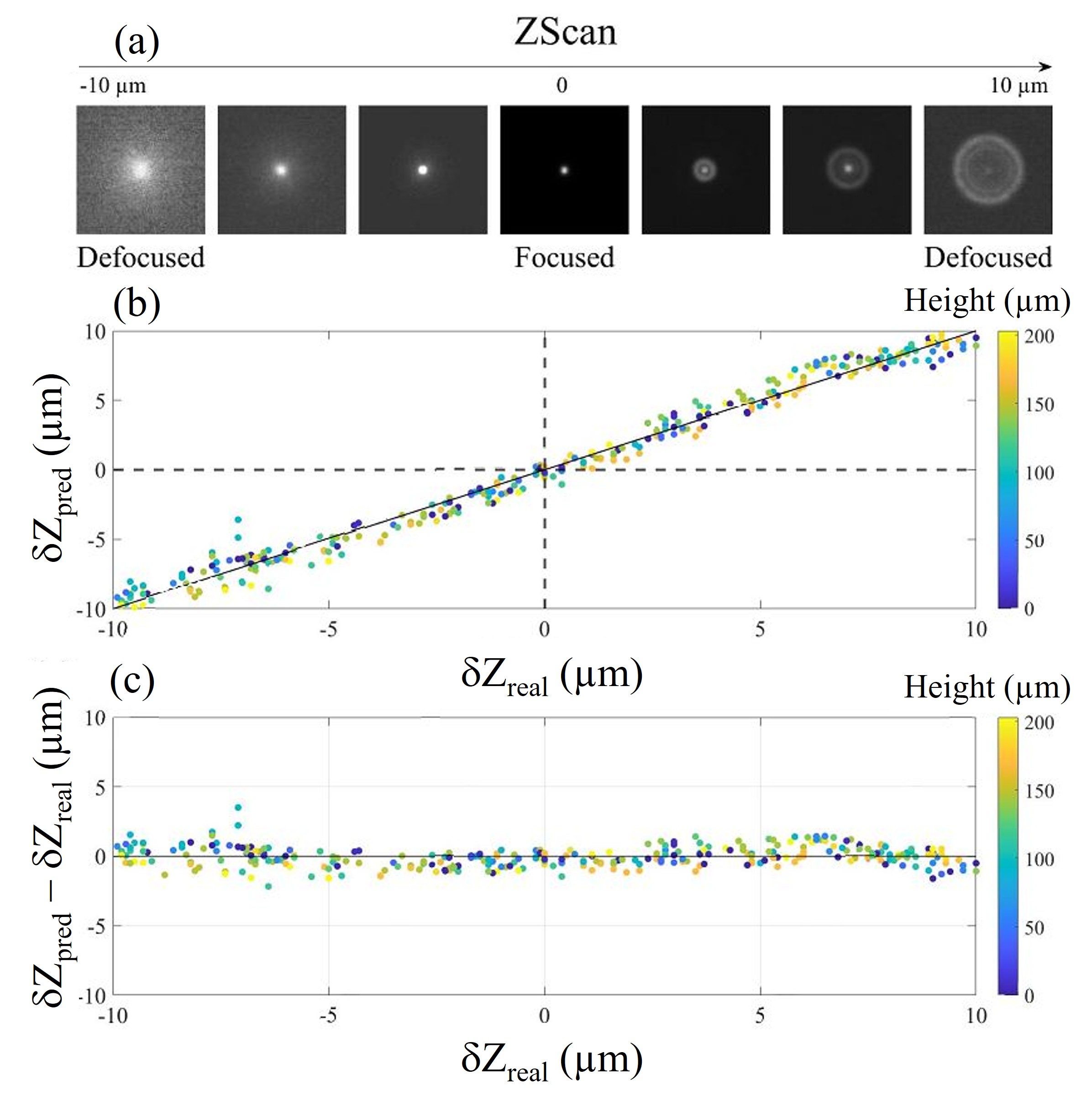}
\caption{
{\bf Axial position determination.}
(a) A ZScan : Stack of object images taken every 0,1 µm at 80Fps (b) Vertical offset from the focal plane $\Delta zpred$  estimated by the AI algorithm, versus  $\Delta zreal = Z_{real}-z_F$, the real position of the bead with respect to the focal plane for different vertical positions (color code). See corresponding Z-scan in Supplementary Video2 and the other $X$,$Y$ predictions in Fig. S4 in supplementary document. The line  is the expected curve for a perfect prediction. (c) At the bottom are displayed the corresponding errors with respect to a perfect prediction.
}
\label{fig_2} 
\end{figure}

Moreover, the deep learning algorithm provides a linear and well-defined response near the focal plane (see also Supplementary Fig. S4 for the linearity of the predictions along the x, y, and z directions). Fig.~\ref{fig_2}(b) plots the measured $\Delta z$ versus z, showing a linear relationship that matches the theoretical curve. This linearity enables the feedback loop to make accurate and proportional adjustments to the microscope's position, maintaining the particle in focus even as it undergoes rapid or complex movements.

%\clearpage

\subsection{Real-time 3D tracking of fluorescent microorganisms}
\begin{figure}[t!]
%\centering
%\includegraphics [width=8cm]{Figures/fig_3 - Full_Image.JPG}
\includegraphics [width=8.7cm]{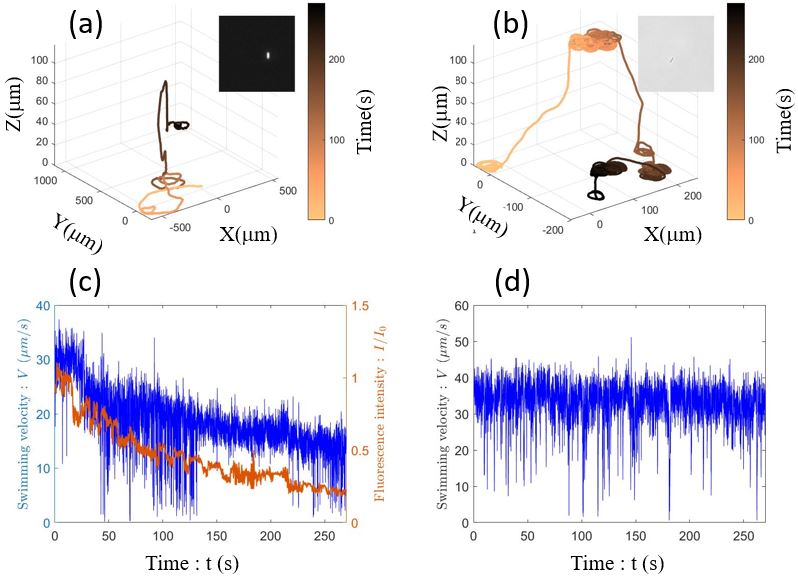}
\caption{\label{fig_3} 
{\bf Long-term tracking of individual bacteria using fluorescence and brightfield microscopy.}
Tracking of a E.coli bacterium with the deep learning-enhanced method (a) 3D trajectory of a fluorescent \textit{E. coli}  obtained using fluorescence microscopy, with an inset showing the bacterium in focus (see Video3a in suppl.videos). (b) In red, corresponding fluorescence intensity decreasing over time due to photo-bleaching (right vertical axis); in blue, decline in swimming velocity over time, indicating photo-damage stemming from a high level extended illumination (left vertical axis). (c) 3D trajectory of an \textit{E. coli} bacterium tracked with brightfield microscopy; inset shows the bacterium in focus see Video3c in suppl.videos).
%(e) Constant fluorescence intensity over time when using brightfield tracking, confirming the absence of photobleaching. 
(d) Stable swimming velocity over time, demonstrating the lack of photo-damage and the ability to track individual bacteria in bright field for long times.
}
\end{figure}

% Tracking of fluorescent {\it Escherichia coli}
% Monitoring swimming velocity and exploration of confinement
% Observation of photodamage effects after extended illumination
To evaluate the capabilities of our deep learning-enhanced Lagrangian tracking method, we first conducted experiments with a E-GFP fluorescent \textit{Escherichia coli} bacterium. The neural network was trained on a stack of fluorescent images acquired around the focal plane at different heights prior to tracking.. See corresponding training Z-scan in supplementary document Video3Fluo and Fig. S6 for predictions.     Fig.~\ref{fig_3}(a) presents the three-dimensional (3D) trajectory of a fluorescent wild-type \textit{E. coli} bacterium %obtained using the standard method, 
with an inset showing the bacterium in focus during the tracking. The trajectory illustrates the bacterium's exploration within the confined space for several hundred seconds. 
A significant issue with fluorescence-based tracking is photo-bleaching. The reduction in fluorescence not only affects the image quality but also poses challenges for the tracking algorithm, which relies on fluorescence signals for feedback. 
Continuous exposure to excitation light causes the fluorescence intensity of the bacterium to diminish over time (bleaching). A second issue is the risk of photo-damage that may affect the bacterium motility when the illumination is carried over extended tracking periods. Prolonged illumination in the blue part of the spectrum (typical excitation range for GFP fluorescence) may lead to photo-damage, adversely affecting the bacterium's motility. To illustrate these issues, we present in Fig.~\ref{fig_3}(b) the result of a  tracking using a high level of blue excitation intensity (twice as much as the one usually used). The photo-bleaching effect (red line) and the steady decrease of swimming velocity (blue line) over time are noticeable. In practice, to overcome these effects, we defined an illumination protocol starting with an excitation at a rather low level of blue light and increase gradually the intensity over time. Nevertheless, the steadiness of the swimming velocity over time could not be maintained for tracking times over $400 $s which may constitute a practical limit of this method when applied to light-sensitive microorganisms.  
%This effect is depicted in Fig.~\ref{fig_2}(b) (red curve), where the fluorescence intensity decreases markedly during the tracking session.  
%
\subsection{Tracking microorganisms using brightfield microscopy}
The versatility of the deep learning-enhanced tracking method permits to overcome these limitations using brightfield microscopy instead of fluorescence microscopy. This is only possible thanks to the deep-learning-enhanced algorithm because  bright-field imaging traditionally lacks the contrast needed for tracking small microorganisms and also can 
be affected by the presence of other objects showing as "dirty" spots. See corresponding training Z-scan in suppl. Video3BrightField and supplementary Fig. S7 for predictions.
Fig.~\ref{fig_3}(c) displays the 3D trajectory of a wild-type \textit{E. coli} bacterium tracked using the deep learning method with bright-field illumination, with an inset showing the bacterium in focus. The trajectory demonstrates smooth and continuous tracking over extended periods, indicating the method's effectiveness without relying on fluorescence. 
The illumination was made using a yellow-color filter to provide a light far from the deleterious illumination  wavelengths that affect bacteria activity.
%
%Interestingly, although fluorescence was not used for tracking, we still measured the bacterium's fluorescence intensity to monitor potential photobleaching. As shown in Fig.~\ref{fig_2}(e), the fluorescence intensity remains constant throughout the tracking session, confirming that photobleaching is effectively eliminated when using brightfield illumination.
%
The swimming velocity of the bacterium remains stable over time, as depicted in Fig.~\ref{fig_3}(d). The absence of a decline in velocity demonstrates that photo-damage is significantly reduced, allowing for prolonged observation of the bacterium's natural motility without adverse effects.
\subsection{Tracking magnetotactic bacteria in bright-field}
Magnetotactic bacteria (MTB) are a group of microorganisms that navigate along magnetic field lines due to the presence of intracellular magnetosomes—organelles containing magnetic iron minerals \cite{blakemore1975magnetotactic}. These bacteria are of significant interest because of their unique biophysical properties \cite{Frankel_2007}. 

However, the need to balance a low $O_2$ environment favorable to magnetosome production and the need to work as much larger $O_2$  concentration favorable to the GFP protein expression, yields a unfitted compromise that may affect in practice either the resulting bacteria motility or the magnetotactic properties . Tentative of adapting fluorescent tracking as described earlier were quite disappointing (see Supplementary information Fig. S8).  
Therefore, the study we present here for MSR-1 is archetypal of many situations where competency to fluorescence is either impossible or difficult to achieve and eventually may affect internal biological properties. In those situation, obtaining reliable 3D tracking for the microorganisms in  bright field (or other non-fluorescent visualization techniques)  is highly needed. 

\begin{figure}[ht!]
%\centering
\includegraphics [width=8.5cm]{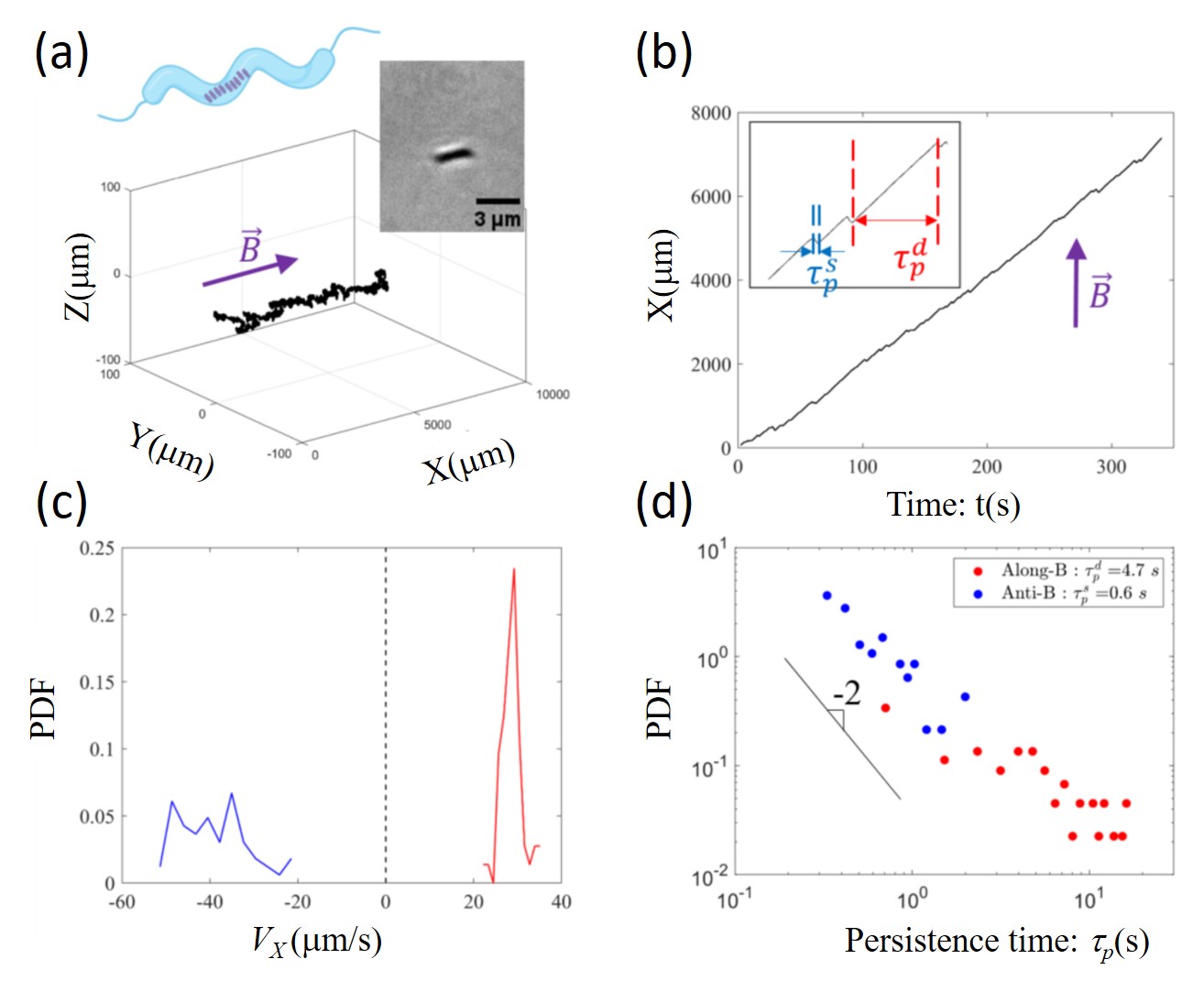}
\caption{\label{fig_4}
{\bf Long-term 3D tracking reveals magnetotactic swimming phenotypes.}
(a) Trajectory of a  magnetotactic bacterium (MRS-1) tracked in 3D under a $ B=1.4$T magnetic field (See  Video4a in suppl. information). Above, image of the bacterium in focus (63X objective) under bright-field illumination. On the right, sketch of the MRS-1 displaying the two amphitricious flagella and the magnetosome (linear assembly of nanoscale $Fe_20_3$ magnetite crystals).  (b) Trace $X(t)$ in the direction of the magnetic field. Inset zoom of the trace showing  sequences of run reversals which compound defines a net motion in the direction of the field (this bacterium is called a "north-seeker"). The inset illustrates a dominant persistent time (along the field)  $\tau_p^d$ and a sub-dominant persistent time (opposite to the field) $\tau_p^s$. (c) Corresponding distributions of the swimming velocities showing a manifest symmetry-breaking characterizing the north-seeker phenotype. (d) Statistics of run-times in the direction of the $\vec{B}$ field (dominant run times in red ) and opposite to it (subdominant run times in blue). 
The solid line is the exponent $\alpha = -2$ as a guide to the eyes illustrating the existence of a fat-tail for these distributions.
}
\end{figure}
Thankfully, our deep learning-enhanced Lagrangian tracking method can  easily be adapted for bright-field microscopy. We trained the NN algorithm on bright field images (see details in Method subchapters \ref{C}1 and in supplementary document Fig. S9 and Video4 ) and 3D tracking in bright-field illumination is now possible.

Fig.~\ref{fig_4}(a) shows an image of a MTB (\textit{Magnetospirillum gryphiswaldense}) \cite{Schuller_1992} in focus under brightfield illumination, along with a schematic representation highlighting its magnetosome chain that provides a magnetic moment of magnitude $ \lVert \vec{M} \rVert \approx 10^{-16}$ J/T \cite{Reufer_2014}. Therefore, by applying a magnetic field non only these biflagellated amphitrichous swimmers will align along the field lines but also will display a propensity to  swim either in the direction of the field (e.g. north-seeker phenotype) or oppositely to it (e.g. south-seeker phenotype). This last property is called magnetotaxis and the conditions of its emergence is still an open and debated issue.
Using the deep-learning-enhanced Lagrangian tracking, we demonstrate the ability  to monitor the bacteria's responses to external magnetic stimuli in real time and over very long time sequences. Fig.~\ref{fig_4}(b) illustrates the  3D trajectories of a MTB under a magnetic field, keeping in average, a swimming direction aligned to the field over more than $8$mm of net displacement (North-seeker). However, looking in detail at the sequence of displacements (see inset) one observes that the bacterium kinematics is quite complex. It displays  sequences of forwards and reversed runs which statistical distribution is at the core of  the  emergence of magnetotaxis. 
%exhibits directed motion along the magnetic field lines, reversing direction when the field orientation is changed. 
The availability of Lagrangian tracking trajectories over long times allows to conduct for the first time, a quantitative analysis of the swimming velocity distribution along a single trajectory. Fig.~\ref{fig_4}(c) displays a clear symmetry breaking for this north-seeker bacterium, whether it swims in the dominant direction of the field (red) or the sub-dominant direction i.e. opposite to it (blue).
%of MTB under a constant magnetic field. Fig.~\ref{fig_3}(c) presents the distribution of swimming velocities, showing a consistent average speed when moving along the magnetic field direction. 
In addition, one can  access the run-time duration for each track and build significant statistics. On Fig.~\ref{fig_4}(d), we show on this example a very striking fat-tail feature for the run-times distribution  that reminds the anomalous motor-switching distribution identified for E.coli rotary motors \cite{Korobkova2004} and attributed to internal fluctuations of a protein responsible for rotation switching of the motor \cite{Tu2005}.
These novel information will be crucial for the understanding of motility characteristics of magnetotactic bacteria and how they respond to a magnetic field.
%
%\clearpage
%
\subsection{Robust tracking in optically complex media}
Tracking microorganisms in optically complex environments, presents significant challenges due to distortions, scattering, and background noise. A traditional tracking method often fails in these conditions because the feedback algorithms cannot reliably distinguish the target organism from the surrounding optical clutter. 
Here, we present two examples of inherently challenging media - (i) crowded bacterial suspensions and (ii) biological mucus- in which our deep learning-enhanced approach provides robust tracking and reveals previously inaccessible information.
\begin{figure}[ht!]
%\centering
    \includegraphics [width=8.5cm]%{Figures/Fig_5.jpg}
{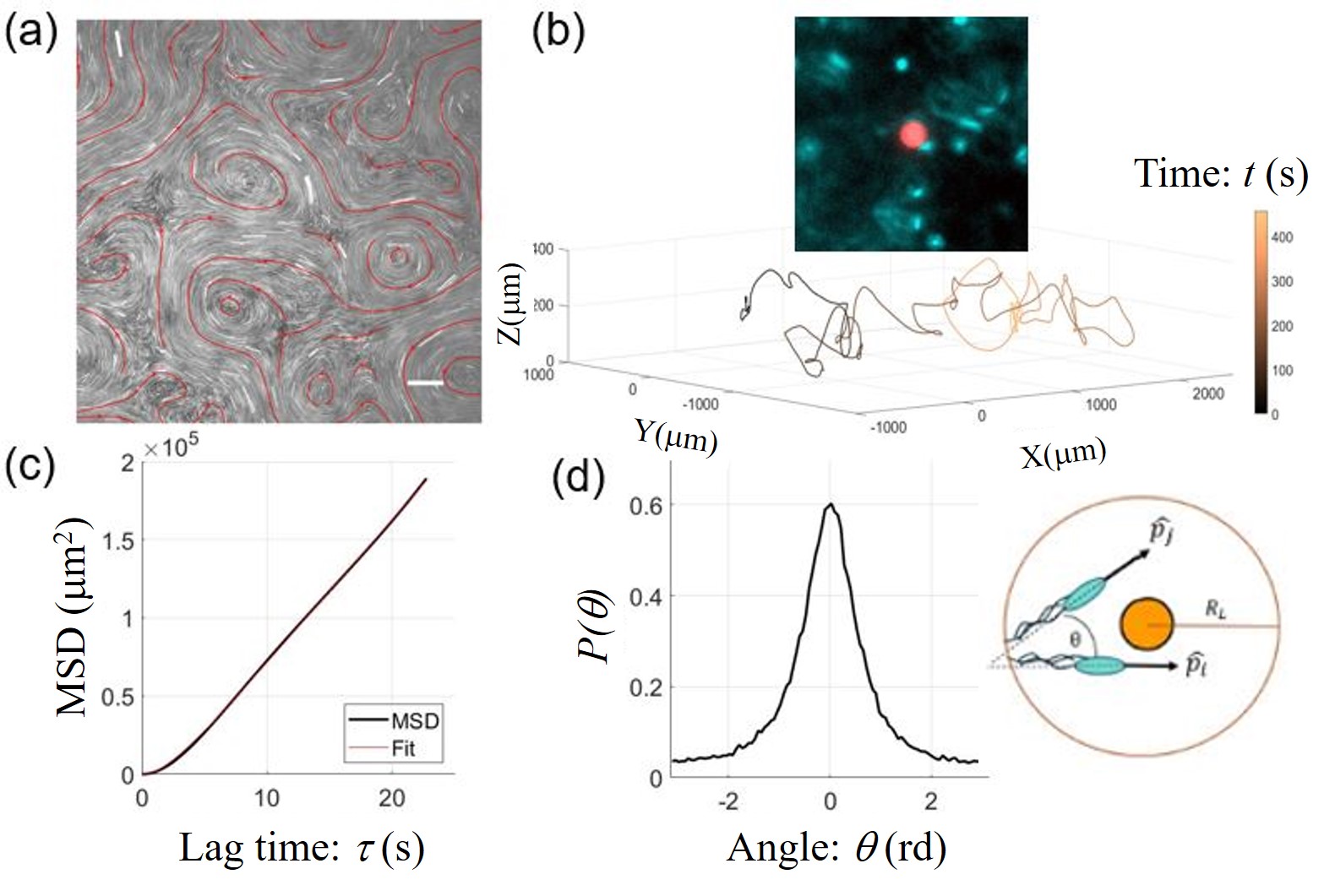}
\caption{\label{fig_5} 
{\bf Transport properties of a passive tracer in a dense bacterial bath.}
(a)  Image of a suspension of motile bacteria in the "active turbulent" regime ,  observed at the mid confinement plane $z = H/2$ with $H = 200 \mu m$ (see Video5a in suppl. information). Flow lines in red are reconstructed by PIV measurement on fluorescent tracers (white tracks represent particle streakline). The white horizontal bar is 100 $\mu m$. (b) 3D trajectory of a $3\mu m$ radius red-fluorescent spherical colloid immersed in the turbulent bath and tracked during  $450$~s. Inset, image of the red fluorescent tracer surrounded by fluorescent bacteria in green (see video 5b in suppl. information). Fluorescent E.coli are about $10 \%$ of the total number of bacteria. Tracking of the colloid  is done with a 2-color technique described in Junot et al.\cite{Junot2022} (c) Long-time transverse MSD of the bead showing a ballistic/diffusive crossover at a time $\tau_c = 2.2 s$. Fit of the MSD (red line) yields a mean ballistic velocity of $45 \mu /s$ and an effective transverse diffusivity $D=475 \mu m^2/s$ . (d) Orientation distribution of relative bacteria velocities around the tracer showing the emergence of a local polar order parameter $P(0) = \langle \cos \theta \rangle = 0.2$.
}
\end{figure}

\subsubsection{Diffusivity of a passive tracer in a turbulent bacterial bath}

Dense suspensions of motile bacteria can exhibit collective behaviors, such as active turbulence, characterized by chaotic flows and large-scale vorticity \cite{alert_2022}. Understanding the transport properties in these active fluids is crucial for applications ranging from mixing enhancement to targeted drug delivery. However, tracking passive tracers within these environments is challenging due to the complex and dynamic optical backgrounds.
We conducted experiments by introducing a fraction of swimming fluorescent bacteria  of $\approx 10\%$ of the total bacteria into a dense suspension of \textit{E. coli} exhibiting active turbulence. Fig.~\ref{fig_5}(a) shows an example of large scale collective motion obtained for a bacterial suspension of volume fraction $\phi= 0.6\%$ inserted in a cylindrical container of radius  $R=0.5$cm and confined vertically to a height $H = 400 \mu$m. Motion of passive fluorescent tracers is analyzed by particle image velocimetry (PIV) \cite{martinez_2020}. The fluorescent tracers  are superposed to the reconstructed flow-lines, showing the large scale vortices characterizing the active turbulent state. 
In similar experimental conditions, we track a colloidal particle using a two-color tracking technique first described by Junot et al.\cite{Junot2022}. Via a beam-splitter, we obtained at the top half of the camera chip  an image of the fluorescent red bead and simultaneously on the down half  an image of the fluorescent bacteria (see the reconstructed image on inset image Fig.~\ref{fig_5}(b)). Using the deep learning algorithm trained specifically for this environment (See corresponding training Z-scan in suppl. Video5), we successfully tracked the 3D trajectories of the passive tracers over extended periods of time that could reach up to one hour.
Fig.~\ref{fig_5}(b) presents a representative trajectory of a tracer particle navigating through the turbulent bacterial suspension. The ability to maintain focus on the tracer despite the surrounding optical noise demonstrates the robustness of our method (see prediction quality details in supplementary Fig. S11) . The transport and mixing  properties of the passive tracer are analyzed on a single track by calculating the mean squared displacement (MSD)  as a function of the time-lag $\tau$ (Fig.~\ref{fig_5}(c)).
%and \ref{fig_4}(d)
The MSD displays a cross-over between a short time ballistic regime and a diffusive regime at long time.
To extract parameters corresponding to  the mean ballistic velocity ($V_0$) and the cross over time $\tau_c$, we fit with a Fürth function and obtain the ballistic velocity $V_0=45 \mu m/s$ and the cross-over time to the diffusive regime $\tau_c=2.2$s, yielding a lateral spreading diffusivity of the tracer : $D=475 \mu m^2/s$ (see details in Chapter Methods \ref{G}).
Furthermore, this technique allows to assess local information corresponding to bacteria around the tracer. Here, we show that we can overcome a recurrent experimental  difficulty due to the fact that in a flow, from a Eulerian measurement, it is not possible to separate swimming velocity and advection by the fluid. Here, provided that the colloid velocity is a good proxy for the local flow velocity,  we are able to investigate the correlations between the swimming velocity directions of the active swimmers and then quantify the level of local  order (polar or nematic) in the turbulent fluid. By removing the local fluid velocity, we directly measure the bacteria swimming orientations, allowing comparisons with theory or simulations, hence providing new insights on collective organization which would not have been possible with previous tracking techniques. In Fig.~\ref{fig_5}(d)), we provide an example of determination of the polar order parameter $<\cos \theta>$ computed locally around the colloid. The average is taken on the bacteria swimming around the probe. 
\subsubsection{Tracking bacteria in mucus}
Mucus is an example of highly heterogeneous and optically complex medium that poses significant challenges for microorganisms tracking. Studying bacterial motility in mucus is essential for understanding infection mechanisms and the development of antimicrobial treatments \cite{McGuckin_2011}. Traditional tracking methods struggle in mucus due to scattering and the non-uniform optical properties of the medium. Therefore, we used the deep learning algorithm trained specifically for this environment (See corresponding training Z-scan Video6 in suppl. information ). In collaboration with a group of medical scientist at Sorbonne University and agronomists of the INRAE in Toulouse, we obtained intestinal porcine mucus extracts similar to those reported in \cite{Mussard_2022,Simpson_2025}. We applied our deep learning-enhanced tracking method to fluorescent \textit{E. coli} bacteria swimming in such a biologically relevant complex fluid \cite{Baillou_2023}. The training methodology is described in  Methods \ref{C}1 and prediction curves in supplementary document Fig. S12. 

Fig.~\ref{fig_6}(a) shows a 3D E.coli track in a porcine mucus extract. From 
such a 3D track one can obtain the  MSD  (see Fig.~\ref{fig_6}(b)) and also the swimming velocities as a function of time (Fig.~\ref{fig_6}(c)). The MSD indicates a diffusive behavior at longer timescales, reflecting the effect of transient trapping and hindered mobility caused by the mucus network (For details see Methods \ref{G}. And indeed, we observed periods where the bacterium's velocity dropped significantly, indicating transient trapping events. Fig.~\ref{fig_6}(d) shows the distribution of transient trapping durations. Such detailed statistical analysis of individual behaviors in complex media was previously unattainable with traditional methods.
%
%%%%%%%%%%%%%
\begin{figure}[ht!]
\centering
\includegraphics [width=\linewidth]{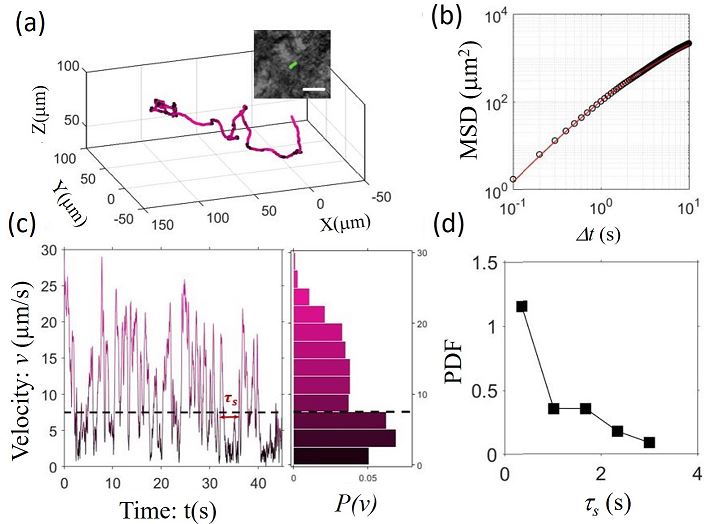}
\caption{\label{fig_6} {\bf Single-bacterium dynamics and transient trapping in intestinal mucus.} 
%Note : for methods : smoothing time used: XY .3s and Z .7s.
(a) 3D trajectory of a fluorescent bacterium tracked in a purified mucus extract from piglet gut for $T\approx 50$s (see Video6a in Suppl. Information). (inset) Snapshot of the bacterium tracked into the optically heterogeneous mucus. Scale bar is $5$µm. (b) Self-averaged MSD of the 3D-trajectory with a ballistic regime at short times followed by a diffusive regime for a time-lag $\tau$ larger than $\tau_c\approx 1$s. The diffusion coefficient is $D\approx50$µm²/s, an order of magnitude lower than what is observed for run and tumble exploration in a motility buffer \cite{Baillou_2025}. In red, we display the Fürth function fit used to extract the ballistic velocity and the correlation time parameters. (c) Instantaneous swimming speed $v(t)$ of the bacterium as a function of time (left panel) revealing periods of motion punctuated by transient stopping times during which the bacterium is almost stuck. Color code is the measurement of the instantaneous speed of the bacterium, increasing from black to purple . The distribution of instantaneous speed (right panel) reflects those two regimes of speed. Below $v=7$µm/s the bacterium is considered as almost stuck .
(d) Histogram of the stopping times for this track, $\tau_{s}$ defined as $ v<7$µm/s, displaying much larger values than the typical tumbling time in motility buffer $\tau_t=0.4$s \cite{Baillou_2025}.
}
\end{figure}
%
%
%\clearpage
%
\section{Discussion}
With our deep learning-enhanced Lagrangian tracking method, we address some inherent limitations of traditional tracking algorithms. This permits us to expand the capabilities of microscopic observation, allowing for the extraction of previously inaccessible data and enabling detailed statistical analysis of individual phenotypical traits in the quantitative analysis of microorganism behavior. This method provides a powerful tool for investigating complex behaviors of microorganisms across a variety of conditions and environments.

A key strength of our approach is its versatility. It adapts seamlessly to different illumination conditions, including both fluorescence and brightfield microscopy. This flexibility is crucial for studying a wide range of microorganisms and particles, especially those that cannot be fluorescently labeled without altering their natural properties. For instance, we successfully tracked magnetotactic bacteria using brightfield microscopy, thereby preserving their magnetotactic characteristics and enabling detailed studies of their response to magnetic fields. This capability opens new avenues for research on organisms previously difficult to study due to labeling constraints.

Our method also demonstrates robustness in optically complex environments, such as dense bacterial suspensions exhibiting active turbulence and biological fluids like mucus. Traditional tracking methods often fail in these conditions due to optical distortions and background noise that interfere with accurate tracking. By using deep learning algorithms trained on environment-specific data, our system accurately tracks microorganisms and particles even in challenging settings. This robustness facilitates studies of transport properties in active fluids and microorganism behavior in realistic environments, which is essential for understanding ecological dynamics and developing medical applications.

The ability to track individual microorganisms over extended periods allows for comprehensive statistical analysis of their behaviors. We extracted distributions of swimming velocities, run times, and trapping durations, providing insights into the motility patterns and interactions of microorganisms in various environments. Such detailed analysis is essential for understanding phenomena like bacterial chemotaxis, magnetotaxis, and the mechanisms by which microorganisms navigate through complex media. For example, studying how bacteria move through mucus can inform the development of treatments targeting infections in mucosal tissues, while understanding transport in active fluids can lead to innovations in mixing processes and drug delivery systems.

Compared to traditional tracking methods, our approach offers increased accuracy and reliability. The deep learning algorithm provides precise position predictions, eliminating ambiguities near the focal plane and ensuring smooth tracking without interruptions. This enhances the quality of the data collected and allows for more accurate modeling of microorganism behavior. Additionally, the method's versatility across imaging modalities accommodates organisms that cannot be fluorescently labeled, broadening the scope of potential research subjects.

Furthermore, the system maintains high performance even in optically challenging media, enabling studies that were previously unattainable. By reducing photobleaching and photodamage—particularly when using brightfield imaging—the method allows for extended observation times, capturing long-term behaviors and rare events critical for comprehensive analysis. The integration of the deep learning model into the feedback loop enables real-time position predictions at rates exceeding 200 Hz, facilitating immediate data analysis and experimental adjustments.

Looking ahead, there are several potential extensions and future directions for this method. Combining our tracking approach with other imaging techniques, such as phase-contrast or differential interference contrast microscopy, could provide additional information about the organisms' morphology and internal structures. Developing adaptive deep learning models that update in real time based on new data could further enhance tracking accuracy in dynamically changing environments.
%Scaling the method to track multiple organisms simultaneously could facilitate studies of interactions within microbial communities and collective behaviors. 
Applying the method within microfluidic devices could enable precise manipulation and control of microorganisms for synthetic biology applications. Additionally, the robustness of the method in complex media positions it well for clinical diagnostics, such as monitoring pathogens in bodily fluids, and for environmental studies, like tracking microorganisms in natural water bodies.
%
%\clearpage
%%%%%%%%%%%%%%%%%%%%%%%%%% METHODS  %%%%%%%%
\section{Methods}
\subsection{Experimental setup and imaging systems  \label{A}}
\subsubsection{Lagrangian tracking microscope}
The swimming microorganisms are observed via an epifluorescence Zeiss-Observer inverted microscope that can also be used in bright-field or in phase contrast, depending on the objective and light conditions. 
 The visualization cells are placed on a 3 axis computer-controlled displacement platform from Applied Scientific Instrumentation (ms-2000-flat-topxyz). 
 Micro-organisms are tracked in three dimensions (3D) using an in-house Lagrangian tracking set-up suited to keep the body of a selected bacterium, in focus and in the central part of the  visualization frame. Images (1024 × 1024 pixels)  are  recorded with a Hamamatsu ORCAFlash 4.0, C11440 camera.
 
The previous methodology and in particular the original tracking algorithm, are detailed in Darnige et al. ~\cite{darnige2017lagrangian}. Note that in a recent version, the issue of backlash readjustment of the two horizontal XY mechanical stages was solved using an optical encoder that delivers the real position of the stages. The Z-position is provided by a very accurate piezo positioning device however limited to a vertical 500 $\mu$m exploration range. The user interface is implemented in a Labview software (using Labview 2021, National Instruments) controlling a trigger that synchronizes the camera readout and the stage XYZ positions collection.

The Lagrangian method was successfully implemented to monitor fluorescent {\it E. coli} swimming in quiescent fluids \cite{figueroa20203d}, Poiseuille flows, both in the bulk and at the surfaces~\cite{junot2019swimming,Junot2022} for a frame rate up to 80 images/s. 
\subsubsection{Details on the previous "minimum searching" algorithm}
In the precedent version, the tracking algorithm used to define the vertical repositioning was essentially based on the size of a defocused object image, minimal at the focal plane. In spite of a non symmetrical size-defocusing relation, the detection of a minimum is inherently problematic when comes to infer accurately the sign of the feedback motion to the Z-positioning stage. For this algorithm, it was essentially the dynamical response, with the ability to counter-react fast if the choice of the displacement it worsening the signal, that allows convergence to the focal plane.  
For large vertical excursions at least in one specific direction, the presence of rings was used  to reorient properly the redirection. However in the other direction rings are not present to unsure a proper feedback motion. These dynamical convergence issues are sources of high frequency fluctuations in position leading to transiently blurred images that can eventually lead to a loss of the tracked object, in particular for longer tracks when the fluorescent signal is bleaching. Therefore\textbf{}, to pursue in this direction a definite effort should be made to improve the Z repositioning part of the tracking algorithm and eventually design a "zero-crossing" method that will supplant the previous "minimum searching" algorithm.  
\subsubsection{Integration of the AI algorithm  }
In the current version of the tracking algorithm a "zero-crossing" method was designed using an AI model trained in Python which principles are described in sections \ref{C} and \ref{D}. We elaborated a Dynamic Link Library (DLL) called by the Labview code to perform the AI predictions.
\subsection{Microorganism cultures and preparation  \label{B}}
%Strains used (e.g., E. coli, magnetotactic bacteria)
We used 3 strains of bacteria:\\
(i) \textit{E.Coli} -  AD62: growth conditions and preparation protocols.  Strain AD62 is pre-cultured overnight (about 14 hours) in Lysogeny Broth Lennox (LB). Bacteria are then cultured for 4 hours by diluting the pre-culture 1:50 in Tryptone Broth (TB : 10g/L tryptone + 5g/L NaCl). OD reaches 0.5.
The resulting solution is then centrifuged (5 minutes at 4590 RCF) and the supernatant medium is replaced by Berg's Motility Buffer (BMB : 3.9 g/L NaCl + 0.1mM EDTA + 25g/L L-Serine + 10mM phosphate buffer 100mM), a minimal medium preventing from bacterial growth. Bacterial solution is then diluted (OD = $1.10^{-4}$) to avoid collision during tracking experiments and mixed 1:1 with Percoll (P1644) so bacteria and the surrounding fluid have the same density (d = 1.06).\\
(ii)\textit{ Magnetotactic bacteria} : growth conditions and preparation protocols. 
MSR-1 bacteria are grown and suspended in Flask  Standard Medium  (FSM) initially elaborated by  Dirk Schuler (for FSM preparation details see Heyen et al. \cite{Heyen2003}). For each new cultivation, 1 ml of fresh FSM is prepared in a 1.5 ml sterilized Eppendorf tube. Then, under the bio-hood, the frozen $-80^{\circ}C$ MSR-1 stock is scratched with a loop and put in an Eppendorf tube inside the incubator for at least one day to allow the frozen MSR-1 to thaw and grow ($T=28^{\circ}C)$. Once the bacteria are active,  100 $\mu$l volume is taken from the Eppendorf tube and transferred into a 15 mL falcon tube containing 9 mL of fresh FSM and put back in the incubator for two days at  $T=28^{\circ}C)$.
For regular daily use, 1mL can be extracted from the previous culture (in the falcon tube) then incubated inside new FSM for one  days (1 ml in  15 mL falcon tube with 9 mL FSM). Then, MSR-1 are ready for a new experiment.\\
(iii) \textit{E.Coli} -  JEK : Growth conditions and preparation protocols. Experiments on collective motion reported in the main text were performed with the non-chemotactic, smooth swimmer strain of \textit{E.coli} JEK1038 (W3110 [lacZY::GFPmut2, cheY::frt], GFP protein not induced for this study). The bacteria were grown in Lysogeny Broth (LB) until the optical density at 600 nanometer (OD) reached $0.6\pm 0.1$. Then, the bacteria were centrifuged and resuspended in a motility buffer (MB) (10 mM K2HPO4, 10 mM KH2PO4, 10 mM sodium lactate, 0.1 mM EDTA, 0.1 mM L-methionine, 0.2 mM L-serine). MB is a minimal medium that prevents cell division but allows bacteria to swim.\\
\subsection{Training data acquisition for deep learning \label{C}}
%
%Procedure for obtaining training datasets
%Acquisition of defocused image stacks at different positions around the focal plane
%Adjustments to optical settings to enhance asymmetry of defocusing patterns
%
Here, we propose a new algorithmic vision to determine the focal plane position and the particle center in real-time. The algorithm here designed is based on the DeepTrack2 Python library \cite{DT, midtvedt2021quantitative, DLCC}.

\subsubsection{Obtaining  the training set samples}
The first step is to train the AI algorithm. In this report we choose to produce training data sets out of real experiments. Note that, in some instances, the data-set stems from synthesized images using the DeepTrack2 Python library \cite{DT,midtvedt2021quantitative}.

A crucial point is to produce  a situation where the object to be tracked will remain immobile during a vertical scanning process of  range $\Delta Z_M$ and  resolution $\delta Z$. We seek  to obtain stacks of images at different positions around the focal plane as well as different heights in the sample in order to capture the vertical variation of the focusing/defocusing patterns. The parameters are to be adapted for the different objects and depend on the magnification and the objective used.\\

For free floating objects, we need to prepare samples and perform scans fast enough such as to minimize the effects of Brownian motion, sedimentation and activity.\\%

- \textit{Training on passive latex beads}. For the training on the $1.1 \mu m$ green fluorescent latex particles displayed in Fig.~\ref{fig_1}, we added percoll to insure density matching and scans were performed with the objective $63XW$ every $\delta Z=0.1 \mu m$ over $\Delta Z_M =\pm 10 \mu m$. \\
However, the $3 \mu m$ red-fluorescent polystyrene colloid used in the bacteria turbulence experiments (Fig.~\ref{fig_5}), we could not use this technique because it is not possible to keep an immobile object in the turbulent bath. So we designed a proxy technique by fixing the colloid to the top of the chamber, made with PDMS. The thickness $H$ of the chamber was varied between 105 µm and 420 µm, while the volume fraction of the bacterial bath was fixed at $\phi=1.6\%$. Scans were performed with a 40X EC Plan Neofluar objective, every $\delta Z =0.5 \mu m$ and  $\delta Z_M = \pm 50 \mu m$ (see supplementary document Fig. S10).\\

- \textit{Training on bacteria}
For the production of the training data set for bacteria, we managed to suppress their motility and insure non-boyancy conditions in different ways.\\

- For \textit{Ecoli} AD62, in Fig.~\ref{fig_3} we achieve non-buoyant conditions by adding percoll. Non-motility condition was obtained by adding ethanol. Finally for experiments in the mucus corresponding to Fig.~\ref{fig_6}, we took the advantage that some bacteria were trapped long enough.\\

- For  \textit{magnetotactic bacteria} MSR-1, corresponding to Fig.~\ref{fig_4}, the density matching and the motility stoppage were both obtained by adding heavy water to the FSM buffer.\\ 

The  training set sample results in a stack of $N$ 1024 × 1024 pixels images coded on a $16$ bit gray scale format. 

Importantly, as a trick to improve the performances of the AI method, we try to enhance the dissymmetry of de-focused images above and below the focal plane. For example the  choice of the objective Apochromat 63XW was  such that we have a correction ring to compensate for the use of glass slides of different thicknesses. We chose to slightly offset the ring in a controlled manner in order to enhance when possible,  the asymmetry. We tuned the ring on $0.19$ mm instead of $0.15$ mm. As a general rule of thumb this "worse is better" strategy has to be adapted for each objective and magnification situations to promote if needed, a proper asymmetry without of course, deteriorating the focal image.\\

Scans were performed at a frequency of $80$ frames per seconds (FPS).
In each situation we obtain a set of about $200$ scans distributed over the relevant vertical range which includes explicitly bacteria positioned at surfaces.\\
Note interestingly that for the red tracer tracked in the collective motion experiments, we used a 40X objective with no technical ring to enhance the de-focusing pattern asymmetry. However, the turbid character of the suspension due to the background bacteria density, seems to participate to the improvement of the symmetry breaking below and above the focal plane.  
\subsubsection{Training set preparation}
An initial step is to determine from each stack (sometimes manually), the image considered as being in focus. This image will correspond to a defocusing index $i_F=0$. The index is then referenced from this image ($i_F> 0$ above the focal plane and  $i_F<0$ below). We crop the stacks in vertical position to keep the images within the defocusing indices $-N_F\leq i_F\leq  N_F$. Here, we use $N_F = 100$ corresponding to an exploration of about $10 \mu$m around the focal plane.

Thereafter, the 16 bit images are converted into a 8 bits images by the transformation :
\begin{equation*}
   I_8 (i,j)={\rm Int}(255 \frac{I_{16}(i,j)-I_{\rm min}} {I_{\rm max}-I_{\rm min}} )
\end{equation*}
where $Int(x)$ is the integer-value function and  $I_{\rm min}$ (resp. $I_{\rm max}$) are the minimal (resp. maximal) values of the 16 bits image value matrix $I_{16}(i,j)$.
The next operation corresponds to a shift of the de-focusing patterns to put them at the image center. 
For this operation, we use in practice a cropping procedure down to a reduced size $L \times L$ that allows (i) to avoid the presence of spurious objects in the periphery (ii) increase the speed of convergence for the center position computation. Here, we used a crop of the whole stack from the centered position of the focused image ($i_F = 0$). The determination of the pattern center coordinate for each image, can be obtained in different ways. We either used Lodestar algorithm of the Deeptrack 2.1 library \cite{Midtvedt2022} or a standard segmentation algorithm to obtain the center of mass. 

After this shift operation for all images in the stack, we obtain $L \times L$  8 bits centered images of the defocusing patterns that will be put in the red channel of the RGB preparation image. Here we use $L=500 $ pixels. 
For the blue channel, we use the single color value:
\begin{equation*}
   I(i,j)={\rm Int}(255 \frac{Z_F} {Z_H} )
\end{equation*}
This information on the vertical position of the focal plane is important as optical aberration and off-focus patterns depend in principle, on this variable and also accounts for the specificity of the surfaces.\\
We then obtain a reference set of training images labeled by the off-focal distance ($i_F$). The central position of the patterns corresponds to translation indices $n_x =0$ and $n_y =0$. In the following, this set will be augmented to account for controlled horizontal shifts and also rotations.

The final training set is made of XY centered images cropped at a reduced size ($L \times L$) (here $L = 500$ pixels) and organized in folders labeled with the defocusing index $i_f$. 
\subsection{Deep learning model training \label{D}}
\subsubsection{Architecture of the neural network}
The training set described in the previous sections was used to train a neural network for predicting the XYZ-coordinate of a defocused particle. The network is a convolutional neural network with 4 max pooling layers, each separated by two convolutional layers. The regression is performed by a dense top consisting of two hidden layers, each having 16 neurons. However, the exact architecture is not expected to be crucial for the performance. 

The very first layer of the network is a non-trainable preprocessing layer, which applies a gaussian filter ($\sigma=10$ pixels) and an average pooling of the input image. This is done to guide the network to focus on large scale features in the image rather than overfitting on irrelevant small scale noise patterns. 
%
%Implementation using DeepTrack 2.1 and TensorFlow
%

\subsubsection{Training-set augmentation}
%description of the different augmentation possibilities including noise and or other particles in the field
To improve the accuracy and robustness of the learning process but also to be able to predict accurately the horizontal motion of the tracked object, the scanned image set is augmented,  first by imposing random rotations and then random translations. This operation is performed automatically using the "Deeptrack" library that produces new sets of images cropped to a value $LX=LY=256$ pixels and labelled  as ($i_F$, $n_x$, with $n_y$) with $-N_M \leq n_x \leq N_M $ and $-N_M \leq n_y \leq N_M $. We use here $N_M = 30$ pixels ($\approx 30 \mu$m). This value can be adjusted to account for the maximal displacement of the object between two images. The training procedure uses a "Tensorflow" model (TensorFlow 2.10, Google) fed dynamically by the augmented data set generated by "Deeptrack".

\subsection{Performance assessment of the AI algorithm  \label{E}}
Once the AI is trained with $80 \%$ of the available images, we reserve the rest $20\%$ for an objective quality test. The prediction performances of the new algorithm is quantified by representing the error between predicted and real positions as a function of the real position.
We present on Fig.~\ref{fig_2}(d) results of the prediction for a GFP fluorescent E.coli suspended in a minimal medium. The training set was achieved using a mixture of minimal medium and percoll ($50\%$ in mass) suited to obtain non-buoyant conditions for the suspended bacteria (see for example \cite{Junot2022}). In this example we trained the AI algorithm using  Lodstar for XY re-centering and for Z with the procedure described previously. We see an excellent predictive power for the algorithm.
 We indeed get a "zero-crossing" method as expected and a determination of the focal position in the central part the prediction with an uncertainty smaller than $0.5 \mu$m.
\subsection{Implementation of real-time tracking algorithm  \label{F}}
%
%Integration of the deep learning model into the tracking system
%Software and hardware synchronization for feedback control
%Use of LabVIEW for user interface and control
%Achieving prediction rates over 200 Hz
%
Using Python based programming for real-time  predictions in order to drive the tracking system, is not possible because of the time latency which has to be very low. Instead, we use the more efficient "Tensorflow" runtime DLL (language C based interface) allowing to dispatch computation on GPUs (Graphical Processing Units). It is called from our Labview tracking software using a home made wrapper DLL. 
Thus, the priory trained models in Python for each training situation, are transported on the Lagrangian tracking machine and called by the Labview tracking program. We are in principle able to reach predictions at a rate larger than $200 Hz$. However, in practice, the current limitations (80Hz for the overall tracking system) come not only from the stream of the image output of the camera but also the response time of the stage control.

%
%
%\subsection{Validation and performance assessment \label{H}}
%Methods for evaluating tracking accuracy and speed
%Comparison metrics with traditional algorithms
%Assessment using Brownian motion of colloidal particles
%Measurement of mean squared displacement (MSD)
\subsection{Computation of the diffusivity $D$}
\label{G}
The diffusion coefficient characterizing the spreading either in plane or in 3D, is extracted from the self-averaged mean square displacement $\Delta \textbf{r}^2 (\Delta t)$ . The position value corresponds either to the trajectory projected on the XY plane $\textbf{r(t)}=(x(t),y(t))$ or the full 3D coordinates $\textbf{r(t)}=(x(t),y(t),z(t))$ : 
\begin{align}
     \left< \Delta \textbf{r}^2 (\Delta t) \right>&=\left< \left[ \textbf{r}(t+\Delta t) - \textbf{r}(t) \right]^2 \right>\\
     &=\frac{1}{T-\Delta t}\int_0^{T-\Delta t} \left[ \textbf{r}(t+\Delta t) - \textbf{r}(t) \right]^2 \,dt
    \label{eq:MSD}
 \end{align}
The mean square displacement is fitted with F{\"u}rth's formula, which depends on two parameters, the ballistic speed and the characteristic cross-over time  $\tau_c$ between ballistic and diffusive regimes: 
 \begin{align}
    f_F(\Delta t) &=2 V_c^2 \tau_c^2 \left( \frac{\Delta t}{\tau_c}- 1 + e^{-\Delta t/\tau_c} \right)\\
    \label{eq:Furth}
    %&\underset{\Delta t \to 0}= V_c^2 \Delta t^2\\
    %&\underset{\Delta t \to \infty}= 2 V_c^2 \tau_c \Delta t = 4D\Delta t
\end{align}
The MSD is fitted on a maximum time lag $\Delta t=T/5$, as a compromise between the statistical convergence of self-averaged quantities and the importance to probe the diffusive regime. Finally, $D = V_c^2 \tau_c / 2d$ where $d$ is the spatial dimension: $d=2$ for lateral spreading (see Fig.\ref{fig_5}) and $d=3$ for the full 3D diffusivity (see Fig.\ref{fig_6}).
%
%Techniques for analyzing trajectories and behaviors
%
%Software tools used for quantitative analysis
%
%Calculation of swimming velocities, persistence times, and 
%
%orientation correlation functions
%
\begin{acknowledgments}
We acknowledge the PushPull (ANR-22-CE30-0038) and BACMAG ( ANR-20-CE30-0034) grants.
EC acknowledges support of the IUF. 
GV acknowledges support from the Horizon Europe ERC Consolidator Grant MAPEI (grant number
101001267), from the Knut and Alice Wallenberg Foun-
dation (grant number 2019.0079), and from the Göran Gustafsson Foundation for Research in Natural Sciences and Medicine.
\end{acknowledgments}
\bibliographystyle{apsrev4-1}
\bibliography{Bibliography}
\end{document}